\documentclass[final,5p,times,twocolumn]{elsarticle}
\pdfoutput=1

\usepackage{times}
\usepackage[caption=false]{subfig}
\usepackage{hepnicenames}
\usepackage{graphicx}
\usepackage[breaklinks]{hyperref}
\usepackage{subfig}

\newcommand{\bbb}{\ensuremath{b\overline{b}}}
\newcommand{\ccb}{\ensuremath{c\overline{c}}}
\newcommand{\ttb}{\ensuremath{t\overline{t}}}
\newcommand{\pt}{\ensuremath{p_{\mathrm T}}}
\newcommand{\ptdijet}{\ensuremath{p_{\mathrm T}^{\mathrm{dijet}}}}
\newcommand{\mdijet}{\ensuremath{m_{\mathrm{dijet}}}}

\newcommand{\madgraph}{\textsc{MadGraph}\xspace}
\newcommand{\powheg}{\textsc{Powheg}\xspace}
\newcommand{\mcatnlo}{\textsc{Mc@nlo}\xspace}
\newcommand{\madgraphamcatnlo}{\madgraph{}\_a\mcatnlo{}\xspace}
\newcommand{\pythia}{\textsc{Pythia}\xspace}
\newcommand{\sherpa}{\textsc{Sherpa}\xspace}

\begin{document}

\begin{frontmatter}



\title{ {\bf Non-resonant Higgs pair production at the LHC in the \bbb\bbb\ final state} }


\author{David Wardrope, Eric Jansen, Nikos Konstantinidis, Ben Cooper, Rebecca Falla, Nurfikri Norjoharuddeen}

\address{Department of Physics and Astronomy, University College London, Gower Street, London WC1E 6BT, United Kingdom}

\begin{abstract}
We present a particle-level study of the Standard Model non-resonant Higgs-pair production process  in the \bbb\bbb\ final state, 
at the Large Hadron Collider at $\sqrt{s}=14$\,TeV. Each Higgs boson is reconstructed from a pair of close-by jets formed with 
the anti-$k_t$ jet clustering algorithm, with radius parameter $R=0.4$\,. Given the kinematic properties of the produced Higgs 
bosons, this Higgs reconstruction approach appears to be more suitable than the use of large-radius jets 
that was previously proposed in the literature. We find that the sensitivity for 
observing this final state can be improved significantly when the full set of uncorrelated angular and kinematic variables of the 
$4b$ system is exploited, leading to a statistical significance of 1.8 per experiment with an integrated luminosity of 
3\,ab$^{-1}$. 
\end{abstract}

\end{frontmatter}

\section{Introduction}
\label{sec:intro}

The thorough investigation of the properties of the Higgs boson discovered by ATLAS and CMS 
\cite{Aad:2012tfa,Chatrchyan:2012ufa} is one of the highest priorities in particle physics for the next two decades. A crucial 
property is the trilinear Higgs self-coupling which can be probed by the study of Higgs-pair production. At the Large 
Hadron Collider (LHC), this is considered to be one of the most challenging processes to observe, even with a data set 
corresponding to an integrated luminosity of 3\,ab$^{-1}$, the target for the proposed High Luminosity LHC (HL-LHC) 
programme. Several particle-level studies were published even before the Higgs discovery~\cite{Baur:2003gpa,Baur:2003gp} 
and more have been published since then, assessing the sensitivity of different 
decay channels such as \bbb$\gamma\gamma$, \bbb$\tau\tau$ and 
\bbb$WW$~\cite{Baglio:2012np,Papaefstathiou:2012qe,Goertz:2013kp,Yao:2013ika,Barr:2013tda}. The \bbb\bbb\ final state was examined in Ref.~\cite{Dolan:2012rv}, where it was 
found to have very low sensitivity, and more recently in Ref.~\cite{deLima:2014dta} where the use of a tighter kinematic 
selection and jet substructure techniques appeared to give some improved sensitivity, although that study considered only the 
$4b$ multijet process as background.

In this paper, we extend our previous work on resonant Higgs-pair production in the \bbb\bbb\ final 
state~\cite{Cooper:2013kia}---which inspired the recent ATLAS analysis~\cite{ATLAS-CONF-2014-005}---to the non-resonant 
case, considering all the relevant background processes, namely \bbb\bbb, \bbb\ccb, and \ttb. The $HH\to\bbb\bbb$ final state benefits 
from the high branching fraction of Higgs decaying to \bbb\  (57.5\% in the Standard Model (SM) for $m_H=125.1\,$GeV~\cite{Aad:2015zhl}, 
leading to about one third of the Higgs pairs decaying to \bbb\bbb), but suffers from large backgrounds. However, like the 
previously studied resonant case~\cite{Cooper:2013kia}, the transverse momentum (\pt) of the Higgs bosons in the 
non-resonant process in the SM is relatively high, with the most probable value around 150\,GeV~\cite{deLima:2014dta}. By 
tailoring the event selection to focus on this high-\pt{} regime, where the two Higgs bosons are essentially back-to-back, one has 
the benefits outlined in Ref.~\cite{Cooper:2013kia} for the resonant case. Requiring four $b$-tagged jets, paired into two 
high-\pt{} dijet systems, is a very powerful way to reduce the backgrounds. This is particularly true for the dominant multijet 
background, which has a cross section that falls rapidly with increasing jet and dijet \pt. There is also negligible ambiguity in 
pairing the four $b$-jets to correctly reconstruct the Higgs decays. Finally, due to the high boost, the four jets will have high 
enough transverse momenta for such events to be selected with high efficiency at the first level triggers of ATLAS and CMS, with 
efficient high level triggering possible through online $b$-tagging~\cite{ATLAS-CONF-2014-005}. We note that triggering will be 
a major challenge at the HL-LHC, but the substantial detector and trigger upgrade programmes proposed by the two 
experiments should make it possible to maintain the high trigger efficiencies reported by ATLAS in the 8\,TeV 
run~\cite{ATLAS-CONF-2014-005} in channels that are essential for key measurements at the HL-LHC, such as the Higgs 
trilinear self-coupling.

\section{Simulation of signal and background processes}
\label{sec:mc}
\begin{table*}[hbt]
    \caption{Summary of the event generators used to model the signal and background processes. The quoted $\sigma \times \mathrm{BR}$ in the last column 
    includes the event filtering described in the text, for the \bbb\bbb, \bbb\ccb, and \ttb\ processes.
    \label{tab:generators}}
  \begin{center}
    \begin{tabular}{cccc}
      \hline\noalign{\smallskip}
      Process & Generator & PDF\ set & $\sigma \times \mathrm{BR}$ [pb] \\
      \noalign{\smallskip}\hline\noalign{\smallskip}
      $HH \to \bbb\bbb$ & \madgraph + \pythia & CTEQ6L1 & $1.16 \times 10^{-2}$ \\
      $\bbb\bbb$ & \sherpa & CT10 & $219$ \\
      $\bbb\ccb$ & \sherpa & CT10 & $477$ \\
      $\ttb$ & \powheg + \pythia & CT10 & $212$ \\
      $ZH \to \bbb\bbb$ & \pythia & CTEQ6L1 & $3.56 \times 10^{-2}$ \\
      $\ttb H(\to \bbb)$ & \pythia & CTEQ6L1 & $1.36 \times 10^{-1}$ \\
      $H(\to \bbb)\bbb$ & \madgraphamcatnlo + \pythia & CTEQ6L1 & $4.89 \times 10^{-1}$ \\
      \noalign{\smallskip}\hline
    \end{tabular}
  \end{center}
\end{table*}

Signal and background processes are modelled using simulated Monte Carlo (MC) event samples. The $HH\to\bbb\bbb$ 
sample is produced with a special release \cite{mg5url} of \madgraph 1.5.12 \cite{Alwall:2011uj}, interfaced to \pythia 8.175 \cite{ref:pythia8} for parton 
showering (PS) and hadronization. This \madgraph release simulates gluon-gluon-fusion Higgs boson pair production using the exact form factors for the top triangle and box loops at leading order (LO), taken from \cite{Plehn:1996wb}. The CTEQ6L1~\cite{Stump:2003yu} leading-order (LO) parton-density functions 
(PDF) are used. The signal cross-section is scaled to 11.6\,fb \cite{Frederix:2014hta}. The \ttb{} events are generated with 
\powheg~\cite{Nason:2004rx,Frixione:2007vw} interfaced to \pythia 8.185 and using the CT10~\cite{Lai:2010vv} PDF set. \ttb{} events with W boson decays to electrons and muons, or where both W bosons decay to light-jets are not simulated, since these decays are suppressed by the requirement for four $b$-tagged jets to pass the event selection, as described in Section~\ref{sec:selection}.
The \bbb\bbb{} and \bbb\ccb{} 
backgrounds are generated by \sherpa 2.1.1 \cite{Gleisberg:2008ta}, using the CT10 PDF set. These event 
samples are scaled to their next-to-leading order (NLO) cross-section by applying a $k$-factor of 1.5~\cite{Greiner:2011mp}.
Other multi-jet processes (such as \ccb\ccb\ and \bbb$jj$) were also considered, but found to be negligible compared to the above 
two, once the $b$-tagging requirements are imposed.
The \bbb\bbb{} and \bbb\ccb{} background samples are filtered at parton level, requiring either: at 
least four anti-$k_t$ $R=0.4$ jets~\cite{Cacciari:2008gp} with $\pt>30$\,GeV and $|\eta|<2.7$; or at least two Cambridge-Aachen $R = 1.2$ 
jets~\cite{Dokshitzer:1997in,Wobisch:1998wt}
with $\pt>150$\,GeV and $|\eta| < 2.7$. In addition, we have considered the most relevant single-Higgs production channels to give 
an indication of their contribution in comparison to the signal and the dominant backgrounds listed above. The $H\bbb$ 
background is generated using \madgraphamcatnlo 1.5.12 \cite{Alwall:2014hca} interfaced to \pythia. The $ZH$ and $\ttb H$ 
processes are both generated using \pythia 8.175. The Higgs mass is fixed to 125\,GeV. More details can be 
found in Table~\ref{tab:generators}.

\section{Discussion of the signal topology}
\label{sec:topology}


\begin{figure}[hbt]
  \begin{center}
    \includegraphics[width=0.45\textwidth]{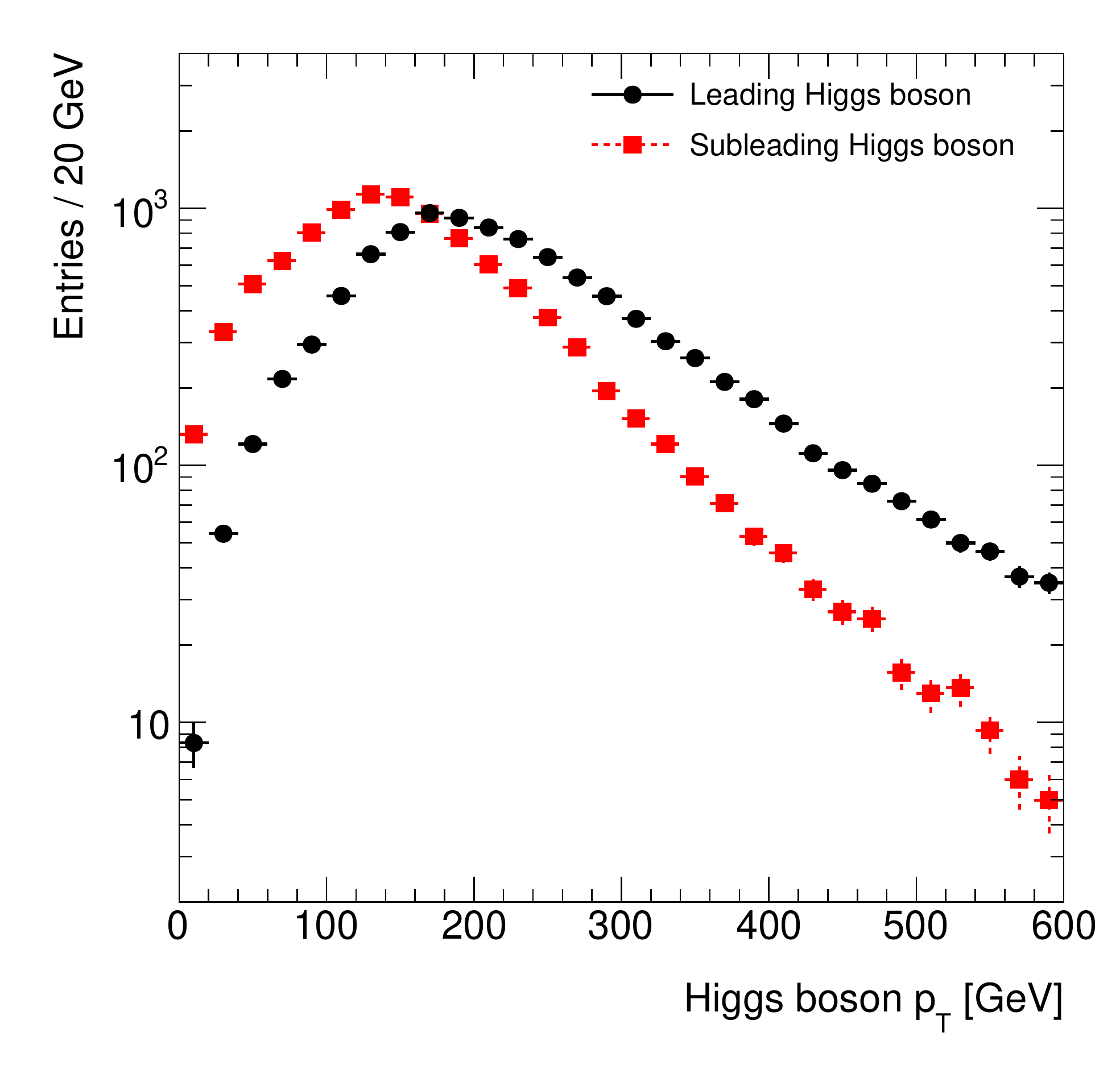}
    \caption {The \pt\ distributions of the leading (circles) and subleading (squares) Higgs bosons in signal events.
      \label{fig:higgspt}} 
  \end{center}
\end{figure}

\begin{figure}[hbt]
  \begin{center}
    \includegraphics[width=0.45\textwidth]{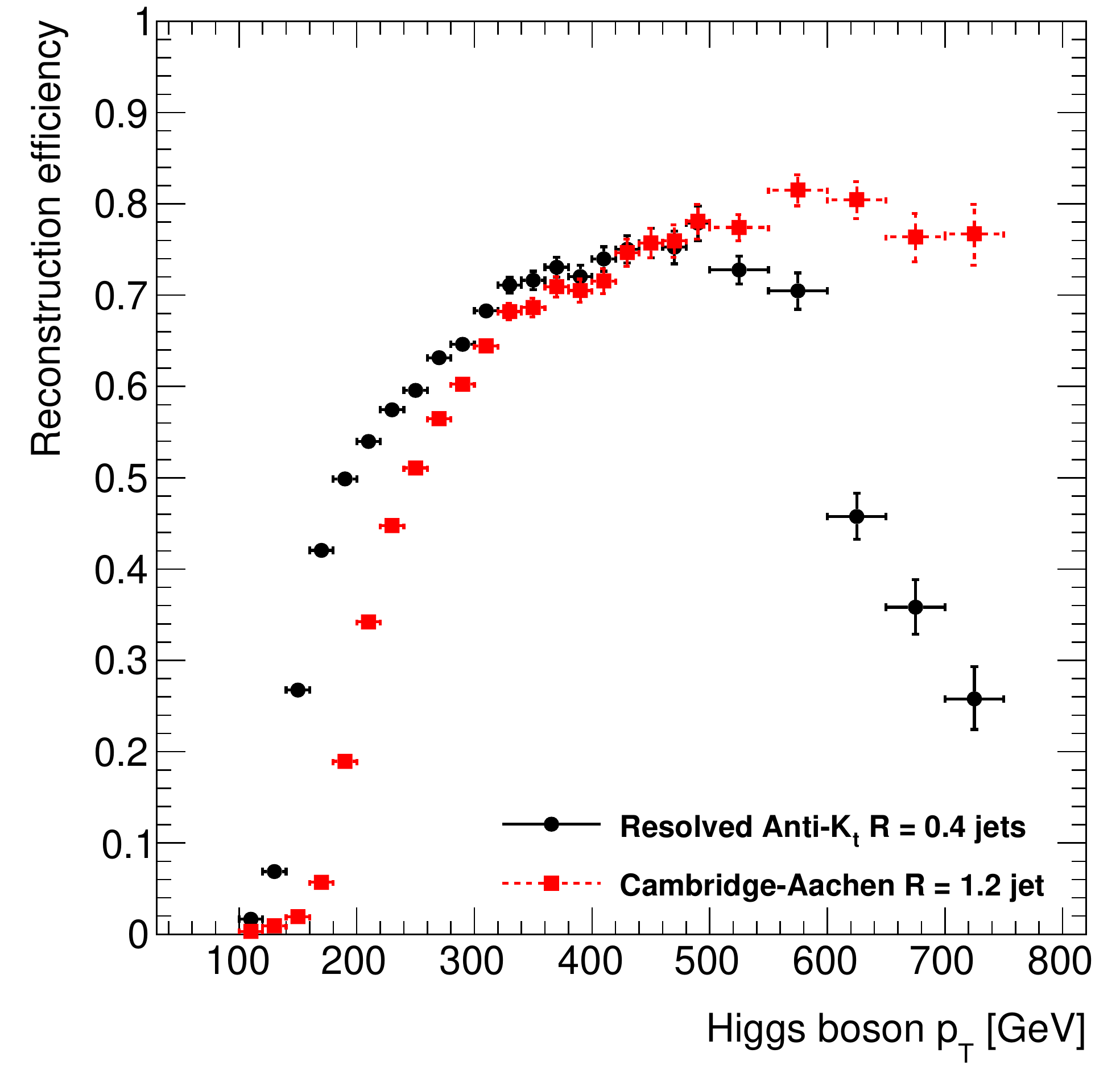}
    \caption {The efficiency for reconstructing correctly the Higgs boson from two anti-$k_t$ jets with $R=0.4$ (circles) 
      or from a single Cambridge-Aachen jet with $R=1.2$ (squares).
      \label{fig:jetalgs}} 
  \end{center}
\end{figure}

Figure~\ref{fig:higgspt} shows the \pt\ distribution of the Higgs bosons in signal events. As mentioned above, in a substantial 
fraction of signal events (36.6\%), both Higgs bosons have $\pt>150$\,GeV. However this drops to 16.6\% (3.6\%) when
requiring both Higgs bosons to have $\pt>200$\,GeV (300\,GeV). 

\begin{figure}[hbt]
  \begin{center}
    \includegraphics[width=0.45\textwidth]{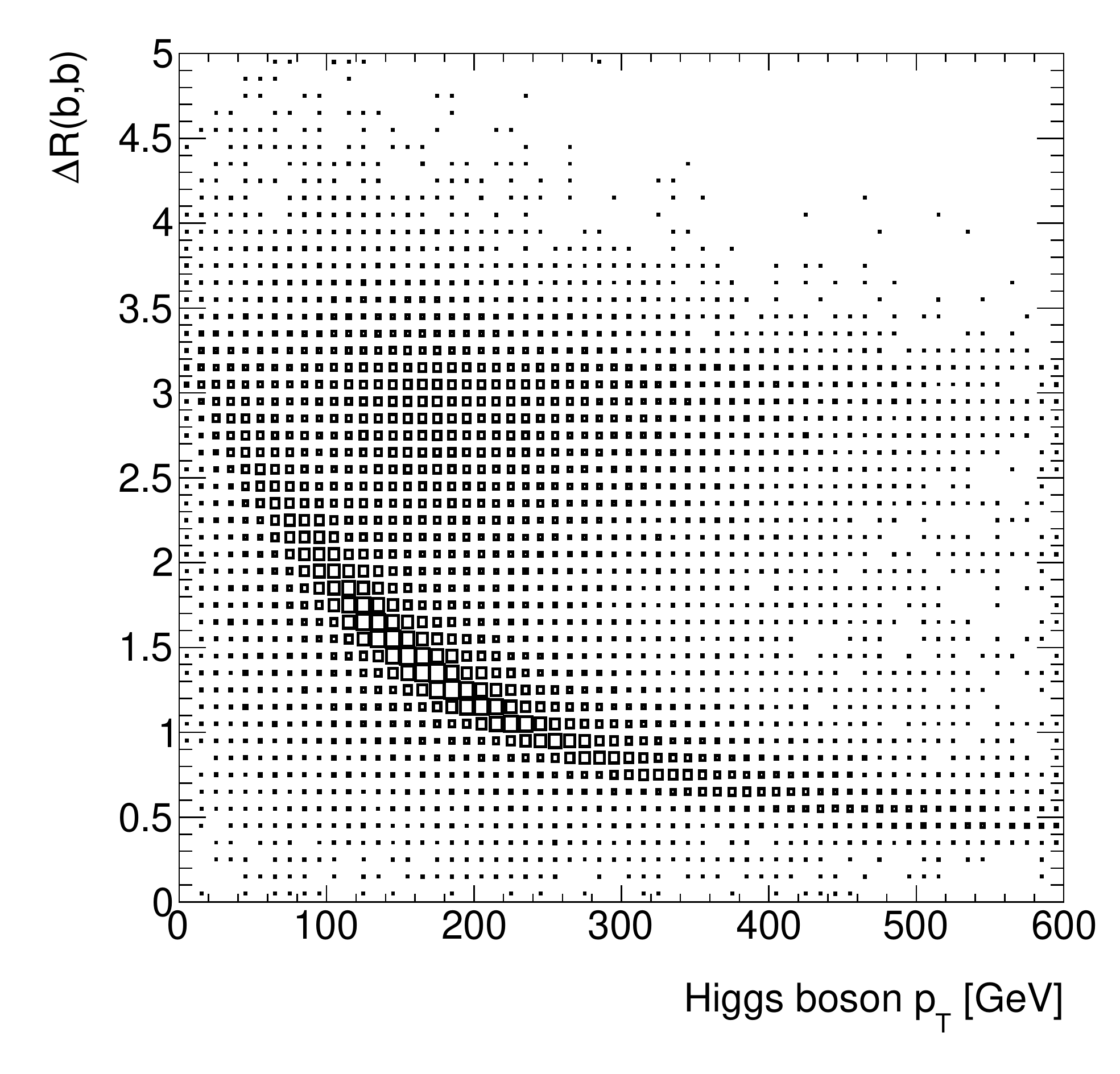}
      \caption {The distance $\Delta R$ between the two $b$-quarks from the Higgs boson decay as a function of the Higgs 
        boson \pt.
          \label{fig:bbdr}} 
  \end{center}
\end{figure}

Figure~\ref{fig:jetalgs} compares the efficiency to reconstruct the Higgs 
boson, as a function of its \pt, using two different techniques: 
(a) combining two anti-$k_t$ jets with $R=0.4$ (hereafter denoted as akt04 jets); and (b) as a single Cambridge-Aachen  
jet with $R=1.2$ (hereafter denoted as ca12 jets). 
In both cases, we use the implementation of the jet clustering algorithms in 
Fastjet~\cite{Cacciari:2011ma} and we include all stable particles in the processing except neutrinos. The efficiency is defined as follows. We take all 
akt04 jets with $\pt>40$\,GeV, and all ca12 jets with $\pt>80$\,GeV containing at least two subjets with $\pt>40$\,GeV (the subjets are formed by reclustering
each ca12 jet using the $k_t$ algorithm~\cite{Catani:1993hr} with $R=0.3$). We then ghost-associate~\cite{Cacciari:2008gn} 
the $b$-quarks from the Higgs decay to all the jets and
subjets. The efficiency for the akt04 reconstruction is defined as the fraction of Higgs decays contained in two akt04 jets 
with angular separation $\Delta R=\sqrt{\Delta\eta^2+\Delta\phi^2}<$1.5\,, where each akt04 jet is associated with a 
different $b$-quark from the Higgs decay. The efficiency
for the ca12 reconstruction is defined as the fraction of Higgs decays contained within a single ca12 jet, with the two $b$-quarks associated to two different 
subjets. From Figure~\ref{fig:jetalgs}, it can be seen that the efficiency of the akt04 approach is higher than the ca12 approach for Higgs \pt\ values 
up to about 400\,GeV. This is not unexpected, given the angular separation of the two $b$-quarks coming from the Higgs 
boson decay as a function of the Higgs boson \pt, as shown in Figure~\ref{fig:bbdr}. At lower Higgs 
\pt, a ca12 jet often cannot capture all the Higgs decay products within its clustering radius. Figure~ \ref{fig:jetalgs} also shows that 
for Higgs boson \pt\ is above 500\,GeV the efficiency of the akt04 approach falls rapidly, but this is not a \pt\ region of interest for the 
non-resonant Higgs-pair production, as can be seen from Figure~\ref{fig:higgspt}.


\section{Event selection}
\label{sec:selection}

The event selection proceeds by requiring at least four $b$-tagged akt04 jets with $\pt>40$\,GeV and $|\eta|<2.5$\,. 
In order to emulate the effect of $b$-tagging in this particle-level study, we adopt the following 
procedure: jets are labelled as $b$-jets, $c$-jets, $\tau$-jets or light jets based on the ancestry of the final-state particles clustered 
into the jet. If a $b$-hadron is found in the history of any of the final-state particles, the jet is labelled a $b$-jet, otherwise if a $c$-hadron is found the jet is 
labelled a $c$-jet. If neither a $b$-hadron nor a $c$-hadron is found, but a $\tau$-lepton is found instead, the jet is labelled a 
$\tau$-jet. All other jets are classified as light jets. We then apply $b$-tagging efficiency weights inspired by the published ATLAS
and CMS $b$-tagging performance~\cite{Chatrchyan:2012jua,ATLAS-CONF-2012-097}: 70\% for $b$-labelled jets, 20\% for 
$c$-labelled and $\tau$-labelled jets (i.e. a rejection factor of 5) and 1\% for light-labelled jets (i.e. rejection factor of 100).
All jets in the event are ordered by $b$-tagging weight and subsequently by \pt. The leading four jets are then used to form dijets, 
requiring $\ptdijet>150$\,GeV, $85<\mdijet<140$\,GeV and $\Delta R<1.5$ between the two jets of the dijet system. If more 
than two dijets satisfy the above criteria, the two which are most back-to-back in the plane transverse to the beam line are 
retained. The two dijets are ordered in $\ptdijet$, and the leading dijet is required to have $100<\mdijet<140$\,GeV, while the 
subleading one must satisfy $85<\mdijet<130$\,GeV \footnote{The mass window for the subleading dijet is at lower masses because often 
in this dijet one of the 
$b$-hadrons has decayed semileptonically, hence the dijet invariant mass shifts to lower values than 125\,GeV and has a larger 
low-mass tail.}. Finally, in order to reject \ttb\ events we use the TMVA framework~\cite{Hocker:2007ht} to train a Boosted 
Decision Tree (BDT) discriminant, $X_{tt}$, using four input variables, two from each dijet system, calculated as follows. We 
search for a third jet with $\Delta R<2$ from the jets of the dijet system, and then calculate: (a) the invariant mass of the 
three-jet system (which would be close to the top mass for a hadronic top quark decay); and (b) the invariant mass of the third 
jet with the least $b$-tagged jet of the dijet system (giving often the $W$ mass in a hadronic top quark decay). Using $X_{tt}$, the 
\ttb\ background is reduced by a factor of $\sim$2.5 for a 10\% reduction in the signal and the multijet background.

After the above selection, the remaining signal cross section is 0.19\,fb, corresponding to about 570 events in 3\,ab$^{-1}$. The 
multijet background cross section is 82\,fb, dominated by \bbb\bbb, and the \ttb\ cross section is 29\,fb, indicating that the 
\ttb{} is a sizeable fraction of the total background. The single-Higgs production $H(\to b\bar{b})$\bbb, $\ttb H$ and $ZH$ 
processes have a combined cross section of 0.33\,fb, comparable to the signal, with the main contribution coming from $\ttb H$. 
Therefore, the signal-to-background ($s/b$) ratio at this point is 0.17\% and the expected statistical significance 
($s/\hspace*{-0.7mm}\sqrt{b}$) for 3\,ab$^{-1}$ is 1.0. Clearly, with such a low $s/b$ ratio, it would be impossible to extract 
any signal sensitivity reliably.

\begin{figure*}[htbp]
  \begin{center}
     \subfloat[$\cos(\Theta^*)$]{\label{fig:thetastar}\includegraphics[width=0.33\textwidth]{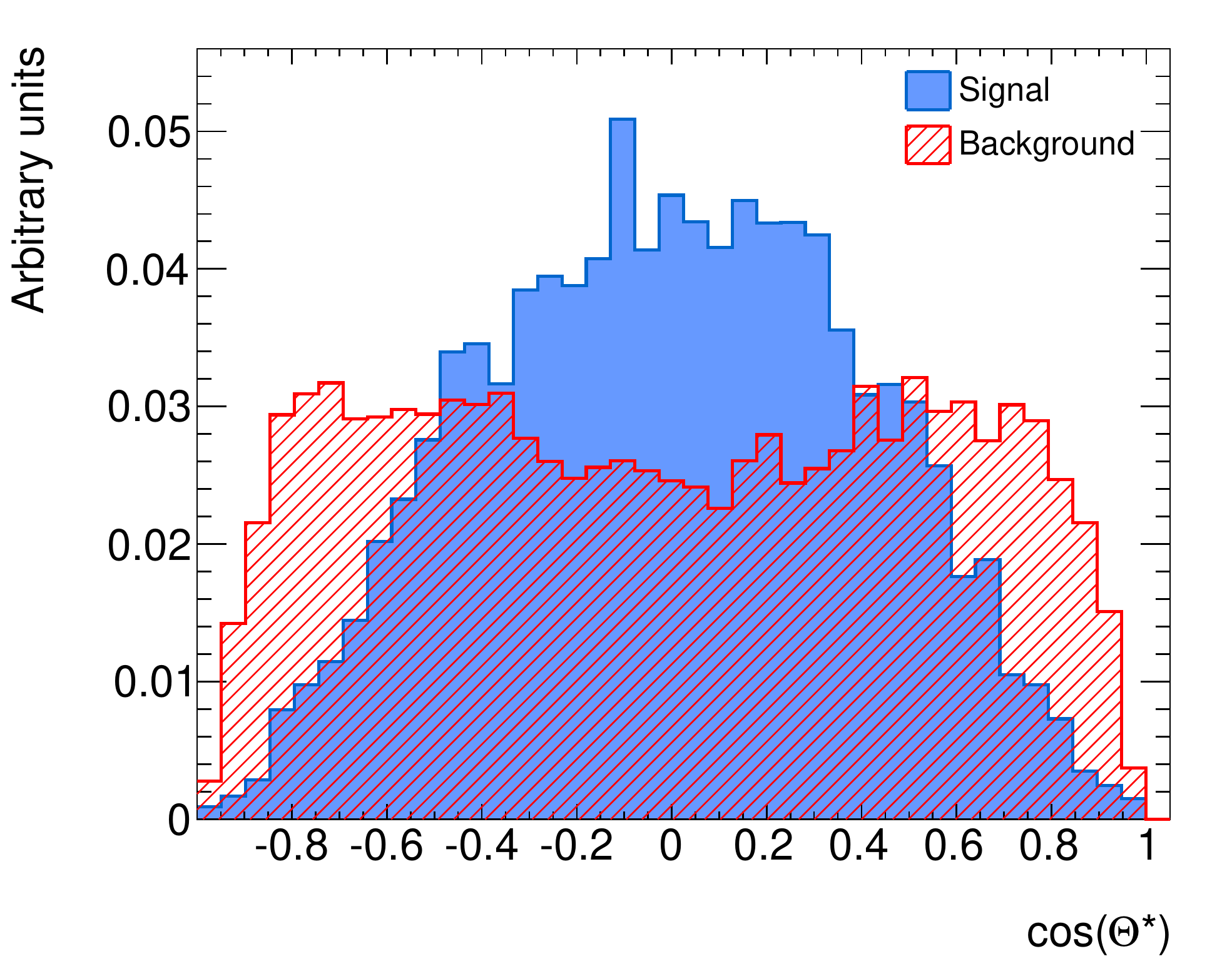}}
     \subfloat[$\cos(\theta_1)$]{\includegraphics[width=0.33\textwidth]{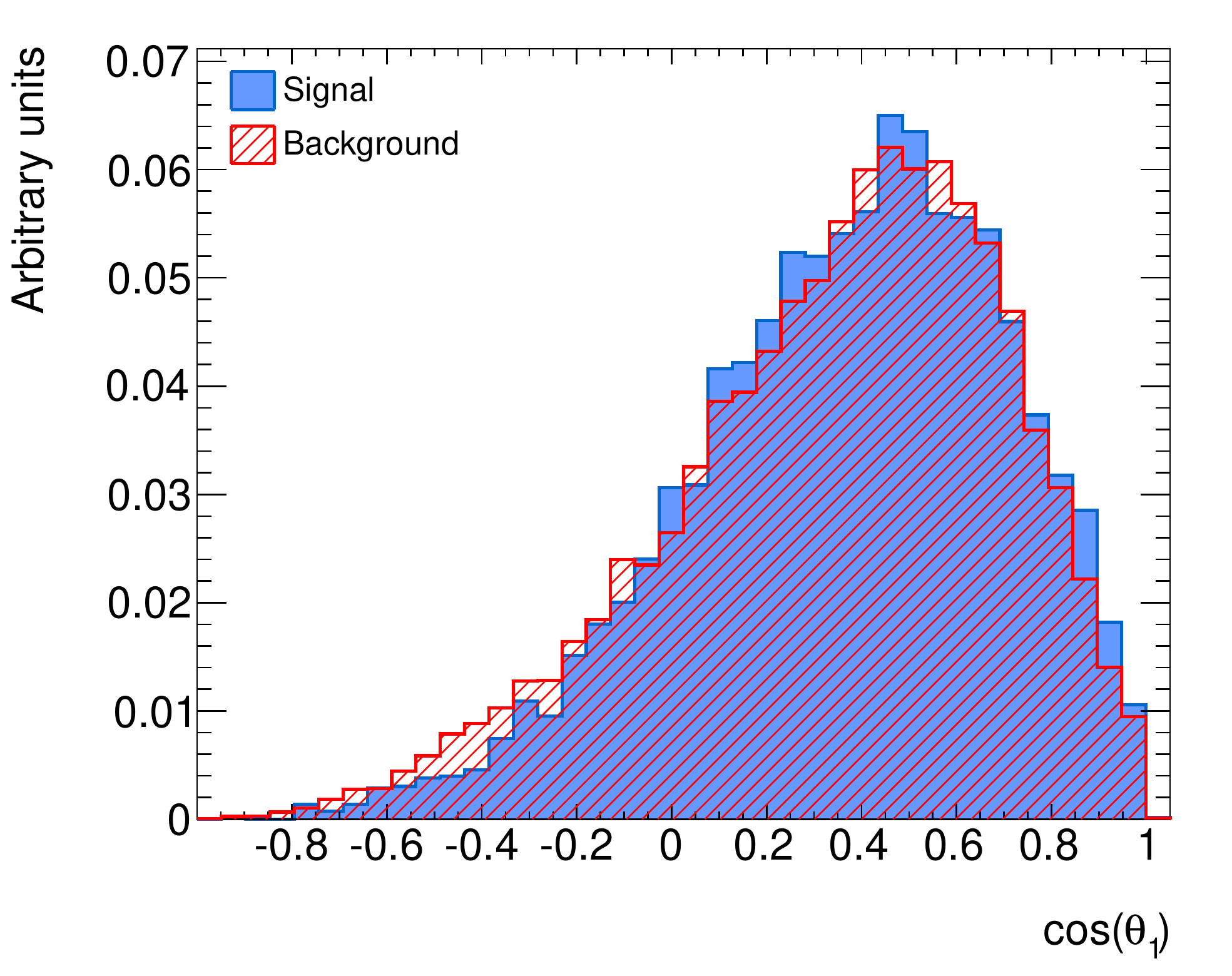}}
     \subfloat[$\cos(\theta_2)$]{\includegraphics[width=0.33\textwidth]{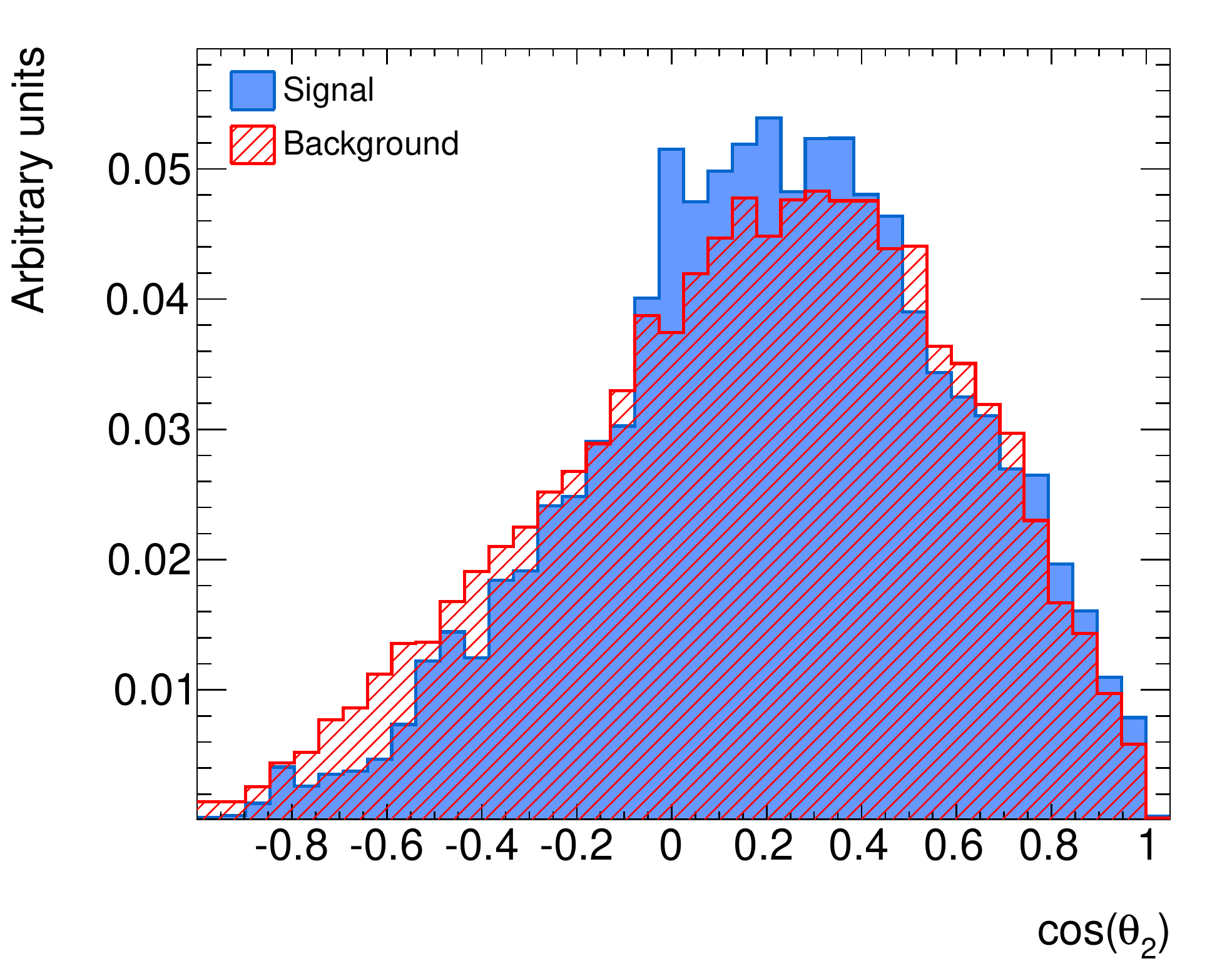}} \\
     \subfloat[$\Phi$]{\includegraphics[width=0.33\textwidth]{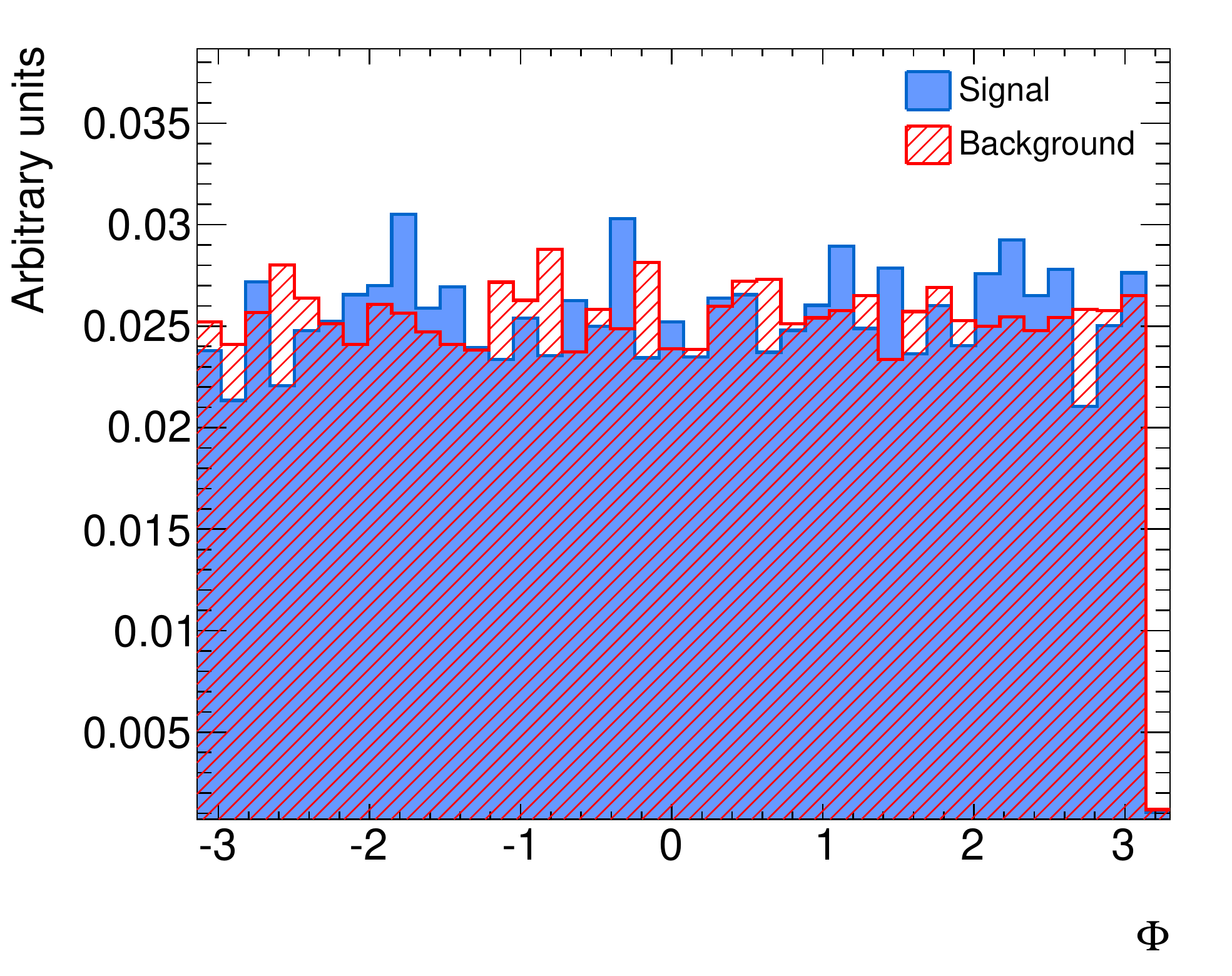}}
     \subfloat[$|\Phi_1| - \frac{\pi}{2}$]{\includegraphics[width=0.33\textwidth]{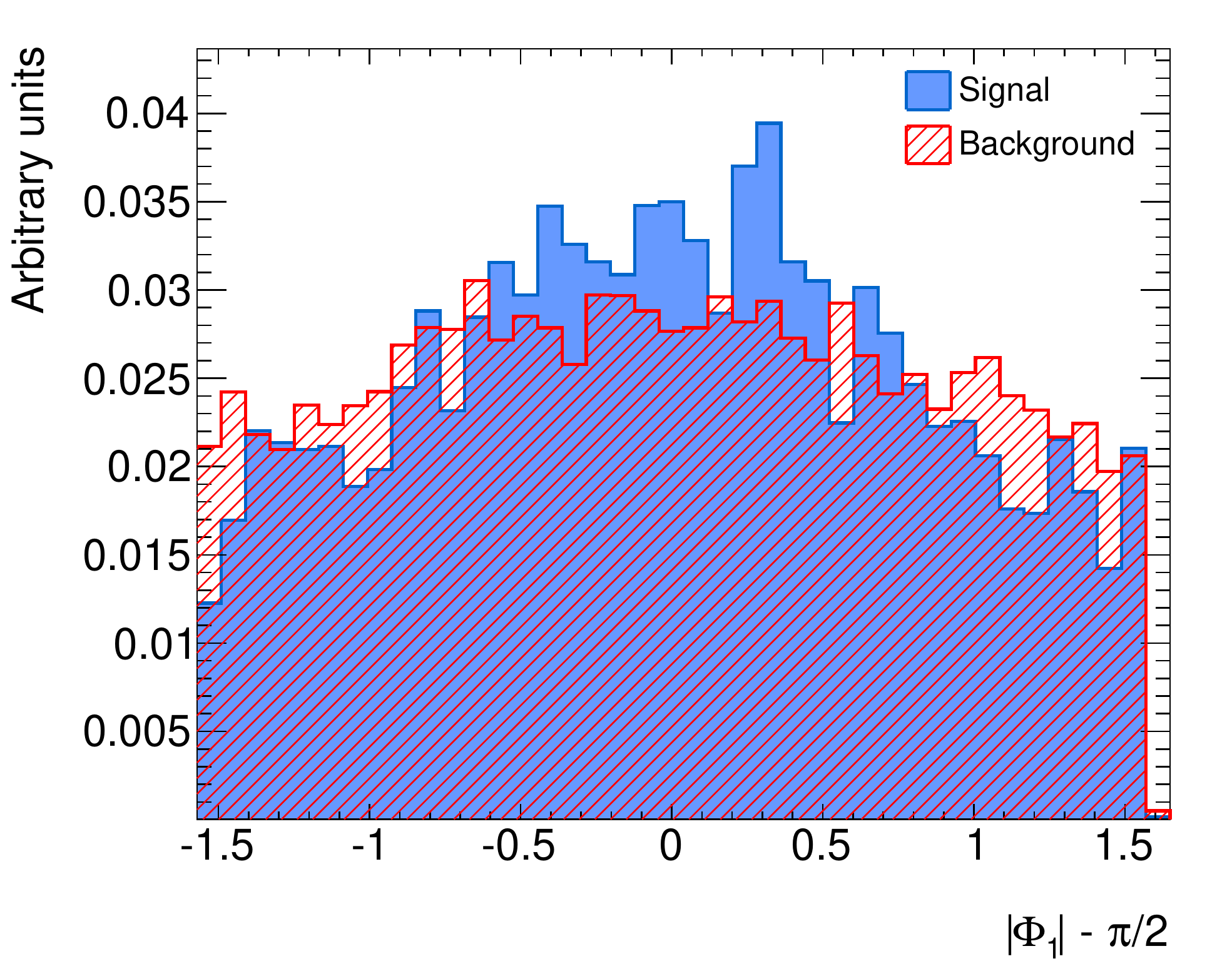}}
     \subfloat[$m_{12}$]{\includegraphics[width=0.33\textwidth]{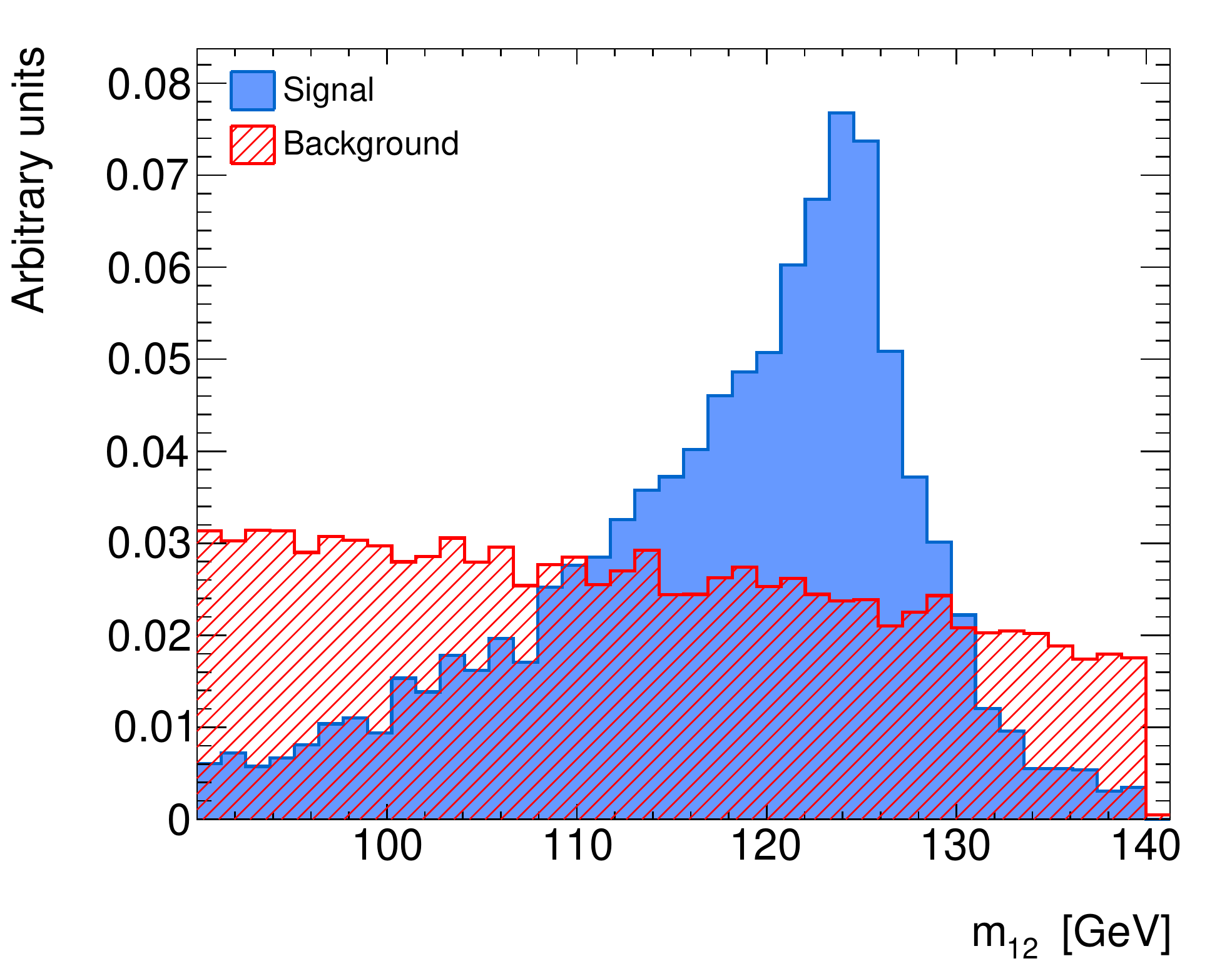}} \\
     \subfloat[$m_{34}$]{\includegraphics[width=0.33\textwidth]{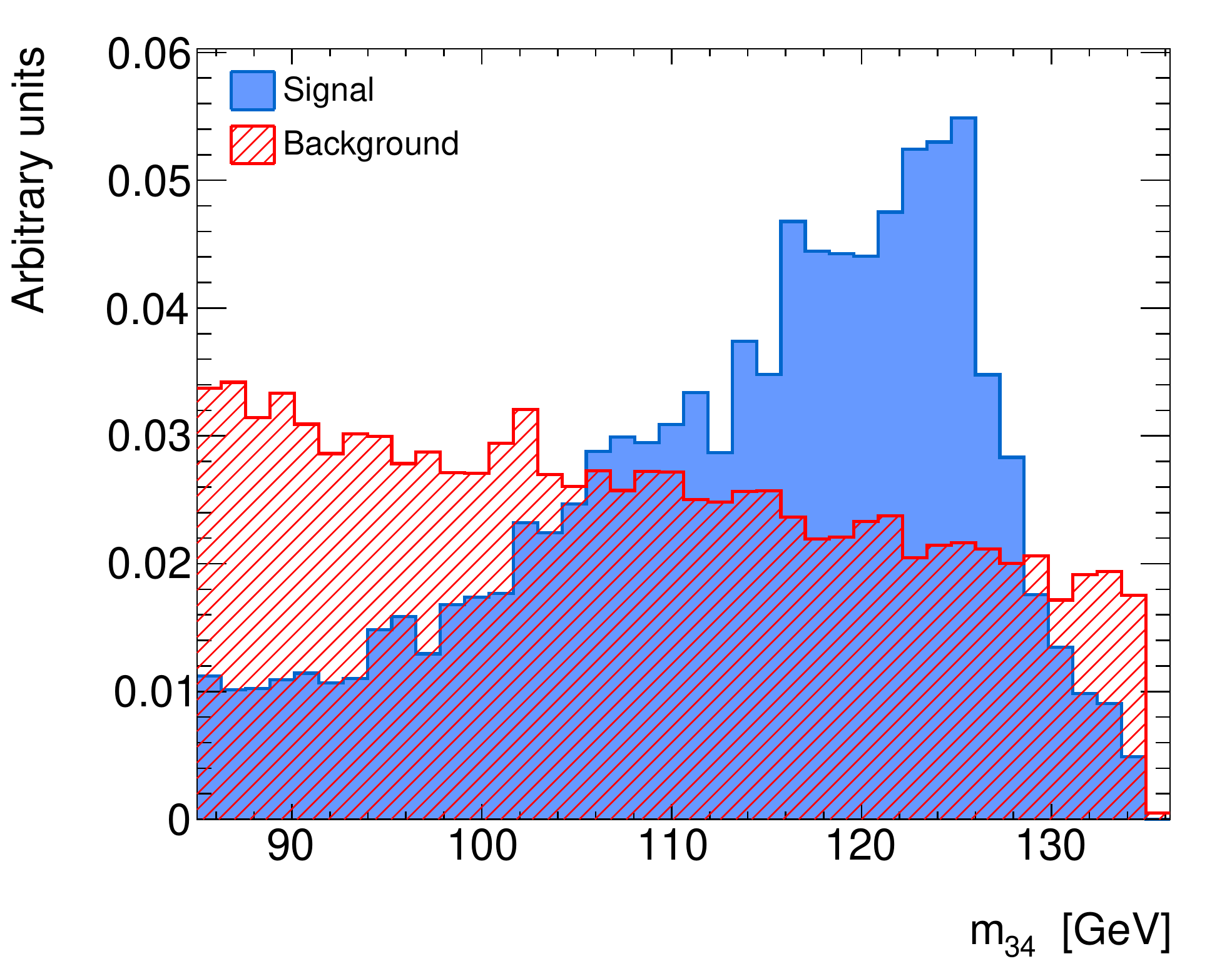}}
     \subfloat[$m_X$]{\includegraphics[width=0.33\textwidth]{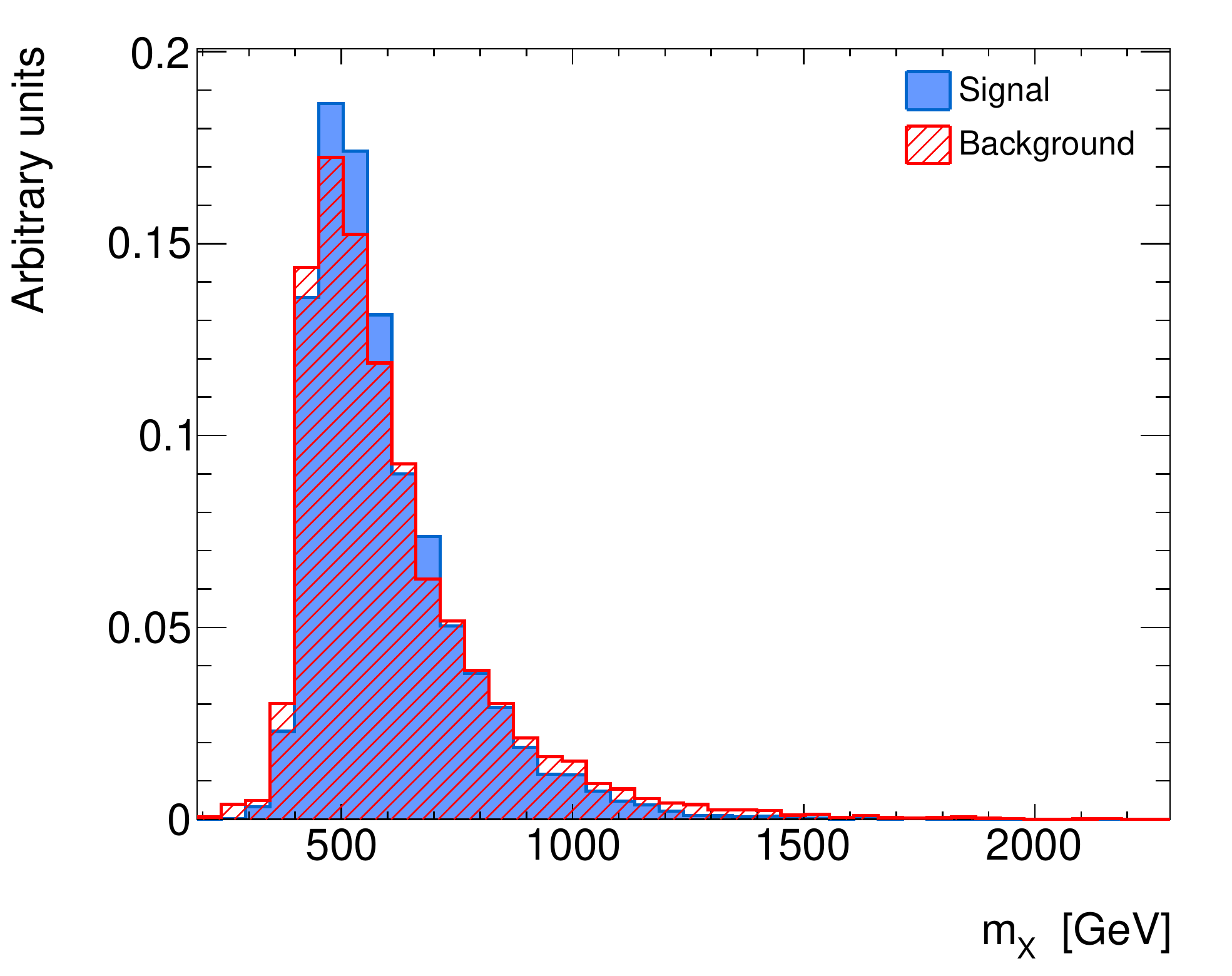}}
     \subfloat[$p_{{\rm T},X}$]{\includegraphics[width=0.33\textwidth]{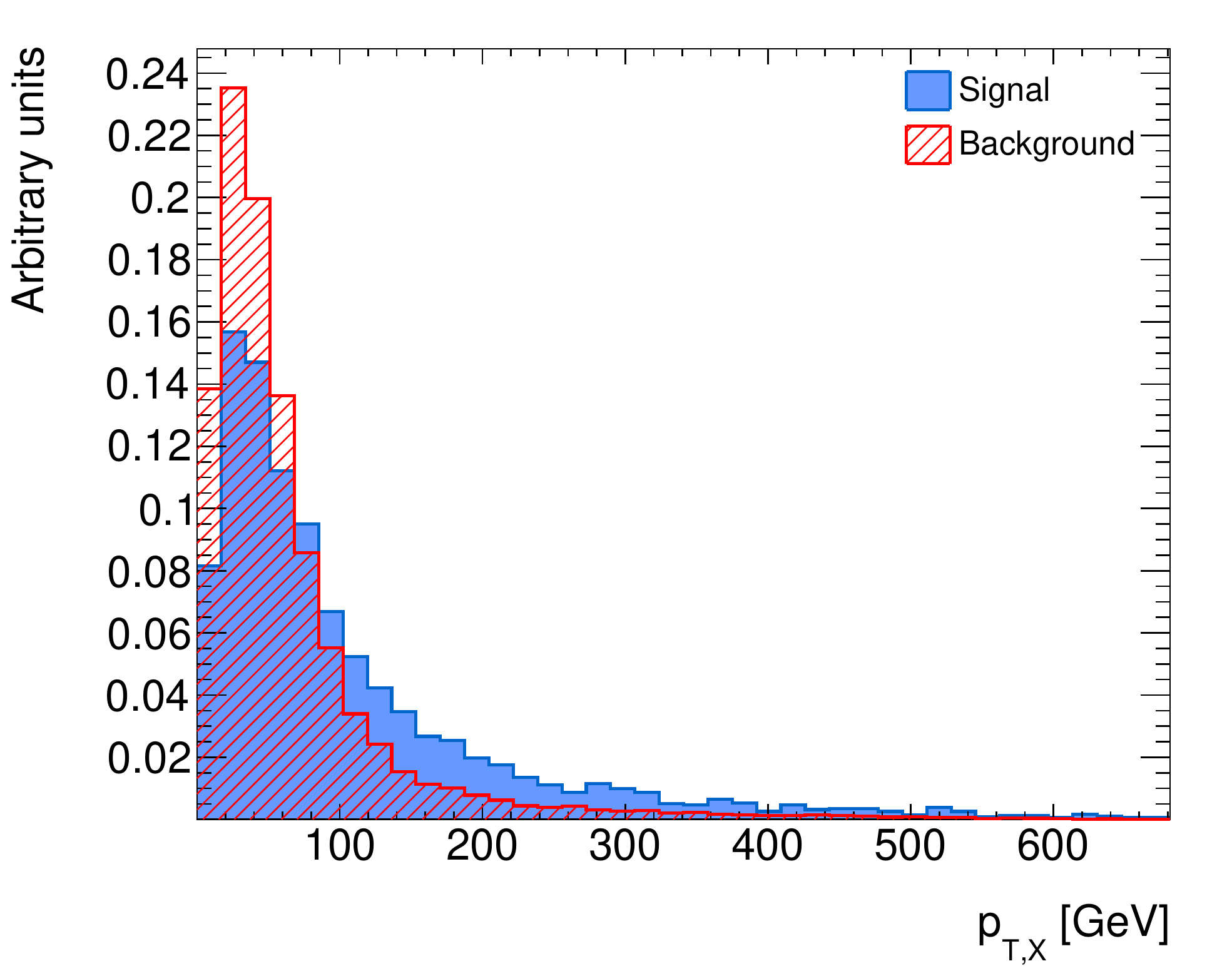}} \\
     \subfloat[$y_X$]{\label{fig:yx}\includegraphics[width=0.33\textwidth]{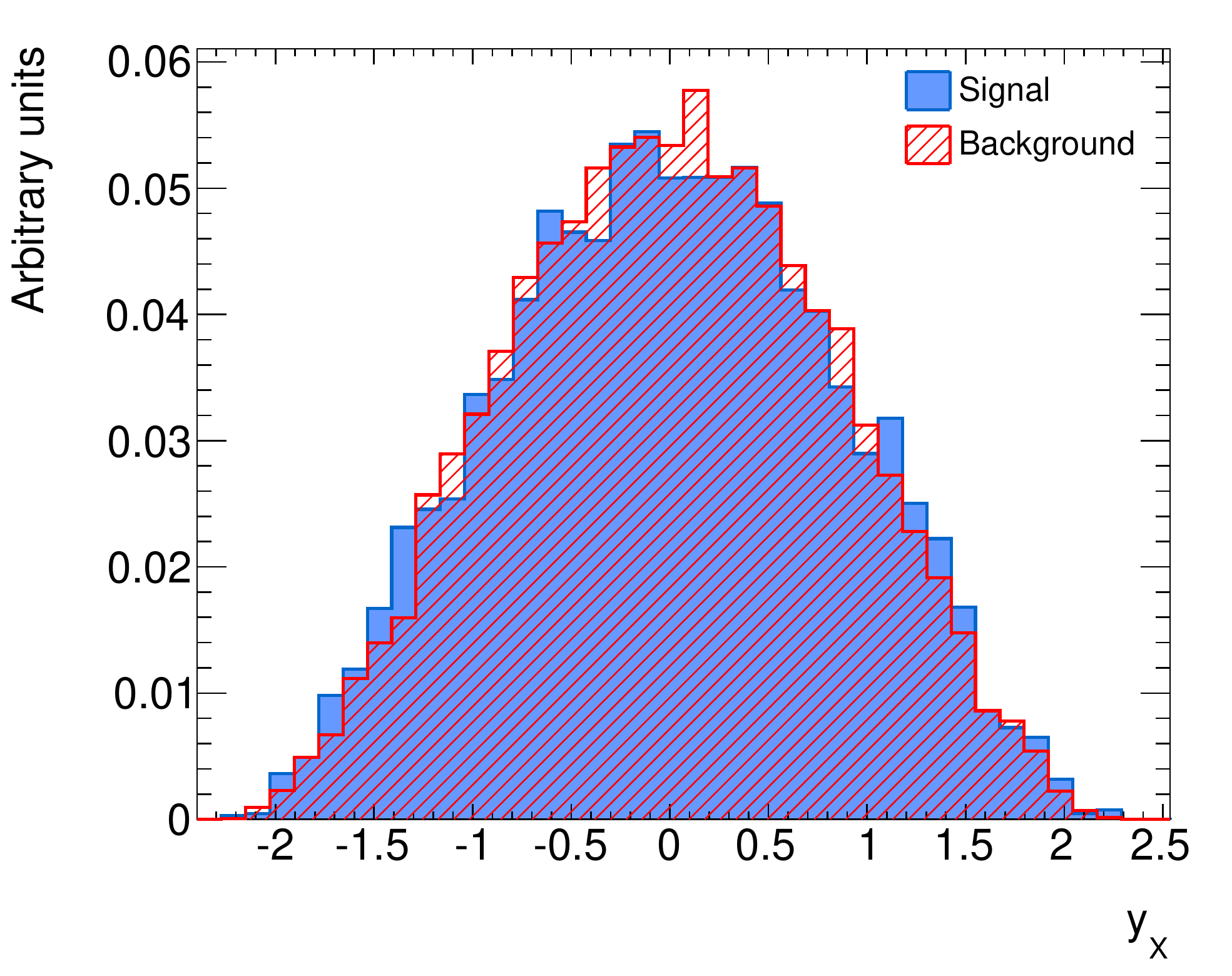}}
     \subfloat[$X_{tt}$]{\label{fig:Xtt}\includegraphics[width=0.33\textwidth]{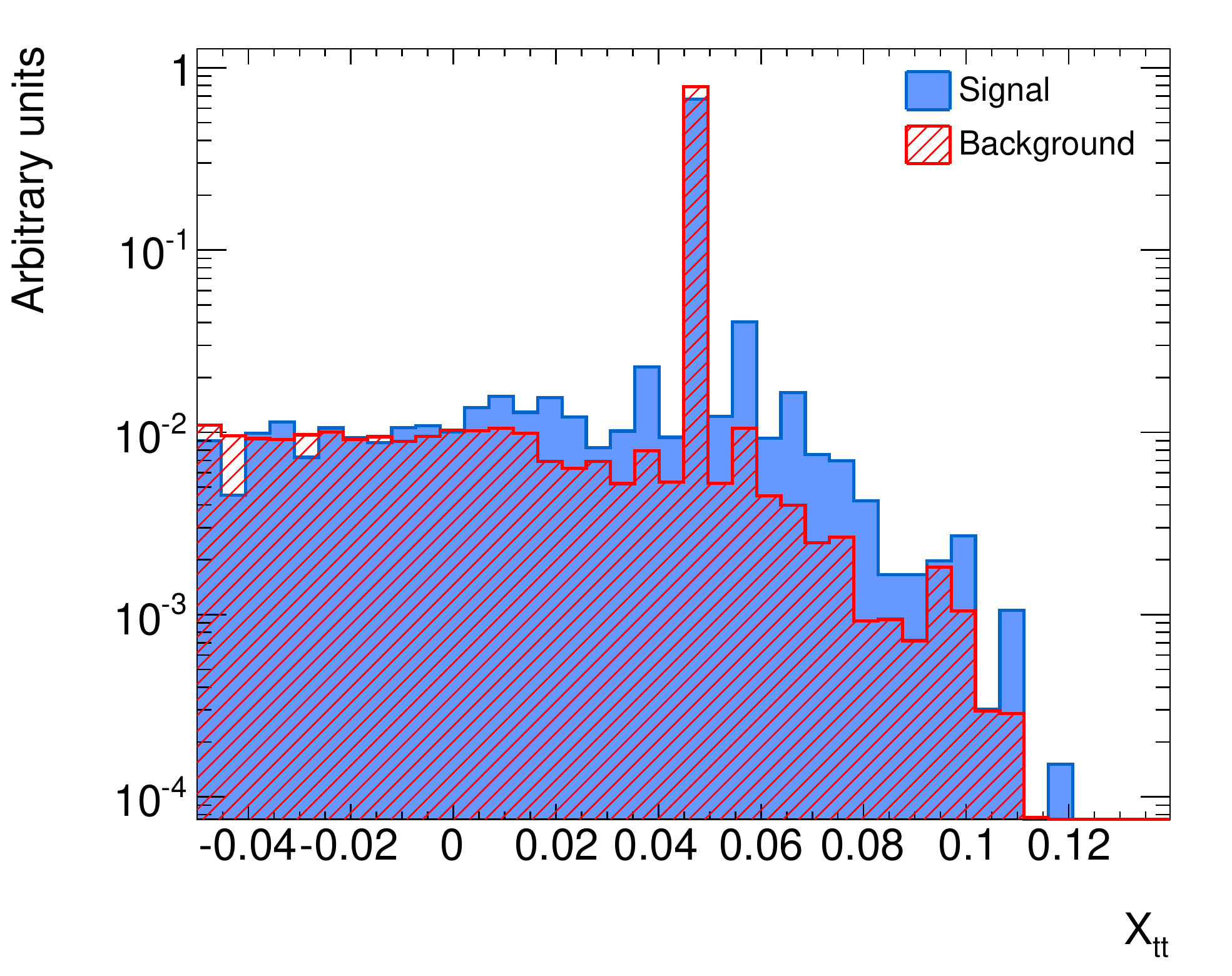}}
     \caption{ Subfigures~\protect\subref{fig:thetastar}-\protect\subref{fig:yx}  show the kinematic and angular variables used 
       to separate the signal and background processes, as described in the text. Subfigure~\protect\subref{fig:Xtt} shows the 
       shape of the \ttb{} discriminant, $X_{tt}$, after the top veto has been applied.
       \label{fig:variables}}
  \end{center}
\end{figure*}

Further to the above selection, any additional kinematic and angular differences between the signal and background can be exploited using the 
following list of largely uncorrelated variables:
\begin{itemize}
  \item{the decay angle of the Higgs bosons in the rest frame of the $4b$ system, $\Theta^*$;}
  \item{the decay angles of the $b$-quarks in the rest frame of the Higgs bosons, $\theta_1$ and $\theta_2$;}
  \item{the angle between the decay planes of the two Higgs bosons, $\Phi$;}
  \item{the angle between one of the above decay planes and the decay plane of the two-Higgs system, $\Phi_1$;}
  \item{the two dijet invariant masses, $m_{12}$ and $m_{34}$;}
  \item{the invariant mass of the $4b$ system, $m_X$;}
  \item{the \pt\ of the $4b$ system, $p_{{\mathrm T},X}$; and}
  \item{the rapidity of the $4b$ system, $y_X$.}
\end{itemize}
These variables have also been proposed~\cite{Gao:2010qx} and used~\cite{Chatrchyan:2013mxa} in the context of the 
$H\to ZZ^*\to 4\ell$ analyses at the LHC. Figure~\ref{fig:variables} shows the distributions of these variables in the signal and 
background after the above event selection. It can be seen, that some of them have little discrimination following the event 
selection, but others show significant differences between the signal and backgrounds.

\begin{figure}[hbt]
  \begin{center}
    \includegraphics[width=0.45\textwidth]{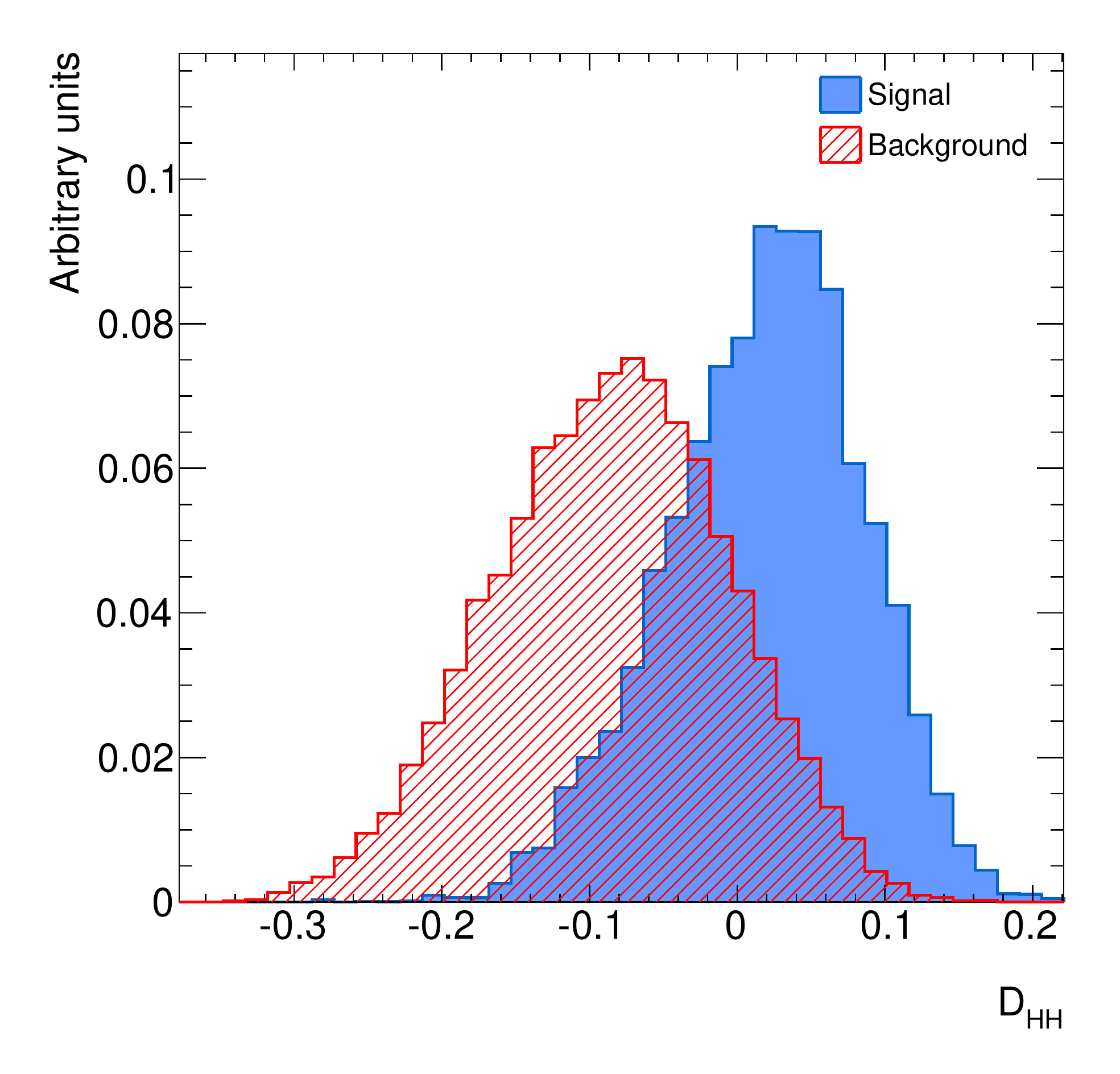}
    \caption {The BDT discriminant ${\cal D}_{HH}$.
      \label{fig:DHH}} 
  \end{center}
\end{figure}

We combine the above variables, together with $X_{tt}$, in a single BDT discriminant, ${\cal D}_{HH}$. The output distributions of this discriminant for signal and background are shown in Figure \ref{fig:DHH}.


\begin{table*}[hbt]
  \caption{Cross sections of the signal and background processes at various steps in the event selection, and the corresponding 
    $s/b$ and $s/\hspace*{-0.7mm}\sqrt{b}$. The last row shows the results for the BDRS analysis described in the text.
    \label{tab:results}}
  \begin{center}
    \begin{tabular}{lccccccc}
      \hline\noalign{\smallskip}
      Requirement & $HH$ [fb] & \bbb\bbb\ [fb] & \bbb\ccb\ [fb] & \ttb\ [fb] & single-$H$ [fb] & $s/b$ & $s/\hspace*{-0.7mm}\sqrt{b}$ (for 3\,ab$^{-1}$) \\
      \noalign{\smallskip}\hline\noalign{\smallskip}
Two dijets           &  0.30 & 513 & 122 & 290 & 2.53 & 3.2$\times 10^{-4}$ &  0.5  \\
$m_H$ windows &  0.21 &  74 &   17 &  73 & 0.65 & 1.3$\times 10^{-3}$ & 0.9  \\
Top veto              &  0.19 &  67 &  15 &  29 & 0.33 & 1.7$\times 10^{-3}$ & 1.0  \\
${\cal D}_{HH}$  &  0.08 & 2.8 &  0.6 &  2.6 & 0.05 & 1.3$\times 10^{-2}$ & 1.8   \\
      \noalign{\smallskip}\hline\noalign{\smallskip}
$\epsilon_{c/\tau-{\mathrm{jet}}}^b=10\%$  & 0.06 & 1.5 & 0.1 & 1.0 & 0.04 & 2.4$\times 10^{-2}$ &  2.1 \\
      \noalign{\smallskip} \hline \hline\noalign{\smallskip}
BDRS analysis    & 0.06 & 11.8 & 1.4 & 6.8 & 0.06 & 3.0$\times 10^{-3}$ &  0.7 \\
     \noalign{\smallskip}\hline
    \end{tabular}
  \end{center}
\end{table*}

\begin{figure}[hbt]
  \begin{center}
    \includegraphics[width=0.45\textwidth]{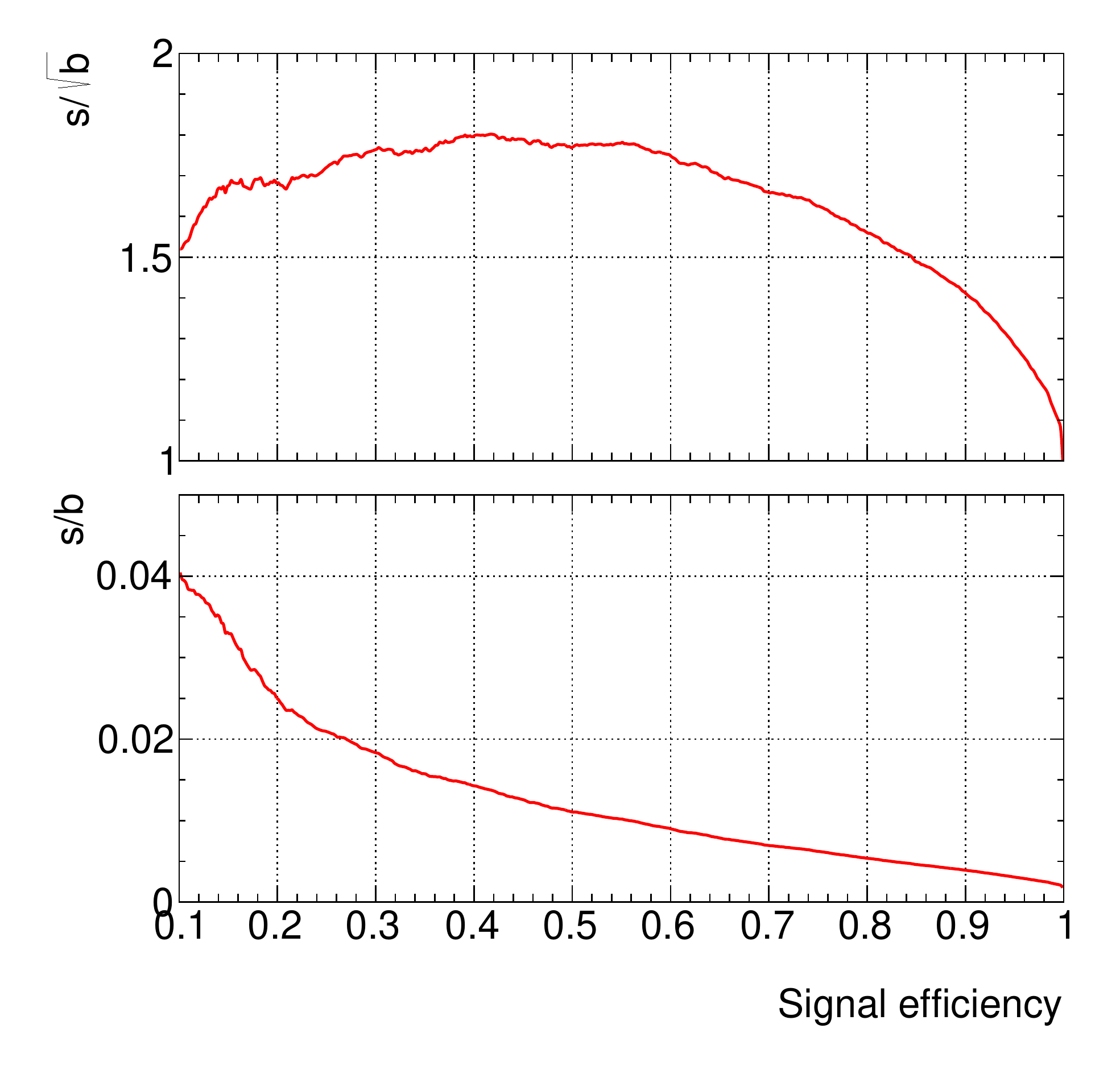}
    \caption {$s/b$ and $s/\hspace*{-0.7mm}\sqrt{b}$ as a function of the relative signal efficiency when varying the cut on 
      ${\cal D}_{HH}$, for an integrated luminosity of 3\,ab$^{-1}$.
      \label{fig:s_over_b}} 
  \end{center}
\end{figure}

\section{Results and discussion}
\label{sec:results}
Figure~\ref{fig:s_over_b} shows $s/b$ and $s/\hspace*{-0.7mm}\sqrt{b}$ for an integrated luminosity 
of 3\,ab$^{-1}$ as a function of the signal efficiency, while varying the cut on ${\cal D}_{HH}$. The highest
statistical significance achieved is 1.8 with $s/b\approx1.3\%$. The signal cross section remaining at this point is 0.08\,fb, 
corresponding to about 240 events with 3\,ab$^{-1}$. The remaining background cross sections are: \bbb\bbb, 2.8\,fb; 
\bbb\ccb, 0.6\,fb; \ttb, 2.6\,fb; and single-Higgs, 0.05\,fb. Figure~\ref{fig:s_over_b} also shows that it is possible to achieve much
higher $s/b$ values for a rather modest decrease in the statistical significance, which may be an important consideration when systematic
uncertainties are also taken into account in the analysis. A summary of all relevant numbers is given in Table~\ref{tab:results}.

These results demonstrate that the \ttb\ and \bbb\ccb\ processes together represent more than half of the total background. 
Most of the remaining \ttb{} background consists of events where the decay products from both $W$'s from the top decays 
include a charm jet or a jet from a hadronic tau decay. This gives additional motivation to improve the charm and tau jet 
rejection of $b$-tagging at the HL-LHC. While the increasing pile-up will make this task challenging, the significantly 
improved pixel tracking detectors proposed for both ATLAS~\cite{ATLAS:1502664} and CMS are likely to provide the necessary 
$b$-tagging performance improvements. In order to demonstrate the potential benefits to this analysis from an improved 
$c/\tau$-jet rejection, we repeated the above study assuming a $b$-tagging efficiency of 10\% for $c/\tau$-labeled jets. On 
doing this, the highest statistical significance obtained is 2.1 at the optimal cut value for ${\cal D}_{HH}$, with $s/b\approx2.4\%$. 

It is worth pointing out that recent theoretical calculations of the SM Higgs-pair production cross section with various 
improvements~\cite{Grigo:2013xya,deFlorian:2013jea,Shao:2013bz} find it is 20-30\% higher than the NLO value used here.
Even if the cross sections of the background processes were increased by a similar factor with more precise calculations, the $s/\sqrt{b}$
would still be 10-15\% better than the result presented above.

In order to have a more direct comparison of the above approach with a selection based on ca12 Higgs reconstruction and jet substructure
techniques in both signal and background, we have applied the BDRS~\cite{Butterworth:2008iy} analysis described in 
Ref.~\cite{deLima:2014dta},  on the signal and background samples listed in 
Table~\ref{tab:generators} \footnote{Using a signal sample generated with Herwig++~\cite{Bahr:2008pv} 
and the same settings as in Ref.~\cite{deLima:2014dta}, we have reproduced exactly the results quoted in that analysis for the signal.}. 
The results of this selection are shown
at the last row of Table~\ref{tab:results}. These results demonstrate that the higher Higgs reconstruction acceptance of the akt04 approach 
combined with all the available angular and kinematic information adds significant sensitivity to the Higgs-pair production analysis.


As this is a particle-level study, it is expected that experimental resolution effects will reduce somewhat the discriminating power 
of the variables used in the above event selection. However, it is worth pointing out that our particle-level predictions in 
Ref.~\cite{Cooper:2013kia} appear to be in broad agreement with the ATLAS result~\cite{ATLAS-CONF-2014-005} that includes 
all the experimental resolution effects and background estimation uncertainties. In addition, there is plenty of scope for further 
optimising the current analysis. Examples of possible avenues to explore for further optimisation include: fitting the distribution 
of ${\cal D}_{HH}$ to extract more information from the data; the use of control regions and data-driven techniques for 
determining the various backgrounds, as in Ref.~\cite{ATLAS-CONF-2014-005}; the use of kinematic fitting techniques to 
improve the angular resolution of the four jets and hence the discriminating power of the angular variables described above; or 
the use of the shape of the $b$-tagging discriminant for each jet, to suppress further the non-4$b$ background events. 

\section{Conclusions}
\label{sec:conclusions}

In SM non-resonant Higgs-pair production at the LHC, the Higgs bosons are mostly produced back-to-back, with relatively large 
\pt. Selecting four $b$-tagged jets and forming two back-to-back pairs, with $\ptdijet>150$\,GeV and $\Delta R < 1.5$ between 
the two jets in each pair, leads to a drastic suppression of all background processes (particularly the dominant multijet 
production) while maintaining a good signal yield. Given the \pt\ spectrum of the Higgs bosons, the use of 
pairs of anti-$k_t$ jets with $R=0.4$ appears to be more suitable for reconstructing each Higgs candidate than the use of single 
Cambridge-Aachen jets with $R=1.2$. 

We further find that exploiting the full kinematic and angular information of the $4b$ system can provide very substantial 
additional improvement in the sensitivity for $HH\to\bbb\bbb$ and the measurement of the Higgs trilinear self-coupling. Our 
particle-level study yields a statistical significance of 1.8 (2.1) per experiment for an integrated luminosity of 3\,ab$^{-1}$, 
assuming a $b$-jet tagging efficiency of 70\% and $c/\tau$-jet $b$-tagging efficiency of 20\% (10\%). While experimental 
systematic uncertainties will tend to reduce 
the sensitivity of the measurement, there is still plenty of scope to optimise the analysis further, hence we expect that the 
sensitivity quoted here should be achievable at the HL-LHC.

\bibliographystyle{atlasnote}
\bibliography{bbbb-pheno}

\end{document}